\documentclass{ckm}             
\usepackage{txfonts}            
\confname{Workshop on the CKM Unitarity Triangle, IPPP Durham, April
  2003}

\newcommand{\be}{\begin{equation}}
\newcommand{\ee}{\end{equation}}
\newcommand{\bea}{\begin{eqnarray}}
\newcommand{\eea}{\end{eqnarray}}
\newcommand{\no}{\nonumber}

\newcommand{\cO}{{\cal O}}

\title{Summary and Overview of Working Group VI: $V_{us}$ and $V_{ud}$}

\author{Gino Isidori}
\address{INFN, Laboratori Nazionali di Frascati, I-00044 Frascati, Italy}

\begin{document}

\begin{abstract}
We briefly review the current status of the determination of $|V_{us}|$ and $|V_{ud}|$,
with particular attention to the latest experimental and theoretical developments
on $|V_{us}|$ since the first CKM Workshop \cite{CKMBook}.
\end{abstract}

\maketitle

\section{Introduction}
Despite the great experimental and theoretical progress in 
semileptonic $b$ decays, at present the most precise 
constraints on the size of CKM matrix elements
are still extracted from the low-energy  
$s\rightarrow u$ and $d\rightarrow u$ 
semileptonic transitions. In a few cases these 
can be described with excellent theoretical accuracy,
and combining the constraints on $|V_{ud}|$ and $|V_{us}|$
we can perform the most stringent test of CKM unitarity.
In particular, the best determination of $|V_{us}|$ is obtained 
from $K \to \pi \ell \nu$ decays ($K_{\ell3}$), 
whereas the two most stringent constraints on $|V_{ud}|$ are
obtained from superallowed Fermi transitions (SFT), i.e.
beta transitions among members of a $J^P=0^+$ 
isotriplet of nuclei, and from the neutron beta decay. 
In addition to these {\em  key modes}, a promising 
and complementary information on $|V_{ud}|$
is extracted from the pion beta decay ($\pi_{e 3}$), 
while significant constraints on $|V_{us}|$
are obtained also from Hyperon and $\tau$ decays.

In all cases the key observation which allow a precise 
extraction of the CKM factors is the non-renormalization 
of the vector current at zero momentum transfer
in the $SU(N)$ limit (or the conservation of the vector current) 
and the Ademollo Gatto theorem \cite{ag}. The latter 
implies that the relevant hadronic form factors
are completely determined up to tiny isospin-breaking
corrections (in the $d\rightarrow u$ case) or $SU(3)$-breaking 
corrections (in the $s\rightarrow u$ case) of second order. 
As a result of this fortunate situation, the accuracy on
$|V_{us}|$ is approaching the 1\% level and the one on 
$|V_{ud}|$ the 0.05\% level. If we make use of the 
unitarity relation 
\be
U_{uu}=|V_{ud}|^2+|V_{us}|^2+|V_{ub}|^2 =1 \ ,
\label{eq:unitarity}
\ee
the present accuracy on $|V_{ud}|$ and $|V_{us}|$ is such 
that the contribution of $|V_{ub}|$ to Eq.~(\ref{eq:unitarity})
can safely be neglected, and the uncertainty of the 
first two terms is comparable. 
In other words, $|V_{ud}|$ and $|V_{us}|$ lead 
to two independent determinations of the Cabibbo angle 
both around the 1\% level, and the unitarity relation 
$U_{uu}=1$ can be tested at the 0.1\% level.

A detailed discussion about the extraction of 
$|V_{us}|$ and $|V_{ud}|$ from the key observables 
mentioned above can be found in Ref.~\cite{CKMBook} 
and will not be repeated here. However, we stress that 
a few significant developments have been achieved since 
the publication 
of Ref.~\cite{CKMBook}:
\begin{itemize}
\item{} The new measurement of ${\rm BR}(K^+_{e 3})$  
by BNL-E865, at the 2\% level of accuracy, 
has been confirmed and finalized \cite{E865}.
\item{} KLOE has announced new preliminary measurements 
of $K^0_{e 3}$ and  $K^0_{\mu 3}$ branching ratios
both at the 2\% level of accuracy \cite{KLOE}.
\item{} Performing a complete analysis 
of $K_{\ell 3}$ decays in CHPT at $\cO(p^6)$, 
Bijnens and Talavera have shown that, 
at this level of accuracy, the amount of 
$SU(3)$-breaking in $f_+(0)$ could be extracted 
in a model-independent way 
from the measurement of slope and curvature 
of the scalar form factor $f_0(t)$ \cite{BT}.
\item{} Cabibbo, Swallow and Winston
have reanalyzed Hyperon semileptonic decays,
showing that theses modes can lead to an 
independent extraction $|V_{us}|$ with a precision 
which is not far from the one presently 
obtained from $K_{\ell 3}$ decays \cite{Cabibbo}.
\end{itemize}
As we shall discuss in the following, these new 
results do not change substantially the overall 
picture presented in Ref.~\cite{CKMBook}, but provide 
a good starting point to reach, within a few years, 
a determination of the Cabibbo angle well below 
the 1\% level.

\section{Status of $|V_{us}|$}
The steps necessary to extract $|V_{us}|$ from each of the  
$K_{\ell 3}$ decay mode can be summarized as follows:
\begin{enumerate}
\item 
  experimental determination of the photon-inclusive 
  decay rate $\Gamma(K\to \pi\ell\nu~+n\gamma)$;
\item 
  experimental determination (or, if not available, theoretical evaluation)
  of the momentum dependence of the two form factor, $f_+(t)$ and 
  $f_0(t)$ (the latter being relevant only for $K_{\mu 3}$ modes);
\item 
  theoretical evaluation of isospin-breaking effects due to 
  both $m_u\not=m_d$ and photonic corrections;
\item 
  theoretical evaluation of the $SU(3)$-breaking correction in $f_+(0)$.
\end{enumerate}
Thanks to the complete $O(p^4, \epsilon p^2)$ analysis 
of isospin breaking corrections in the framework of CHPT
($\epsilon$ stands for both $e^2$ and $m_u-m_d$)
by Cirigliano {\em et al.} \cite{CKNRT},  
the theoretical error due to the step n.~3 is around 0.3\%. 
This means that if we combine the first three steps 
in this list for the four different $K_{\ell 3}$ decay modes,
we should obtain four independent determination of the 
product $f_+(0)|V_{us}|$ affected by a very small
theoretical error. 

The master formula for a combined analysis of this type is:
\bea
&& |V_{us}| \cdot f_{+}^{K^0 \pi^-} (0) = \left[ \frac{192 \pi^3  \Gamma_i}{  G_{F}^2 M_{K_i}^5 C_i^2
 S_{\rm ew} \,  I^0_{i} (\partial f_+, \partial f_0) } \right]^{1/2} \no \\
&& \qquad \times  \frac{1}{1+ \delta^{i}_{SU(2)} +   \delta^{i}_{e^2 p^2} + \frac{1}{2}
\Delta I_{i} (\partial  f_+, \partial f_0) }~,
\label{eq:master}
\eea
where $C_i=1$ ($2^{-1/2}$) for neutral (charged) modes,
$S_{\rm ew}= 1.0232$ denotes the universal short-distance electroweak 
correction factor \cite{sirlin} and 
$I^0_{i} (\partial f_+, \partial f_0)$ the non-radiative phase 
space integral. The small correction terms in the second line 
of Eq.~(\ref{eq:master}) denote the  isospin-breaking effects 
computed in \cite{CKNRT} and reported in Table~\ref{tab:iso-brk}.

\begin{table}[t]
\centering
\begin{tabular}{|c|c|c|c|}
\hline
& $\delta_{SU(2)} $ & $\delta_{e^2 p^2} $ &
$\Delta I (\partial f_+, \partial f_0) $ \\
\hline
$K^{+}_{e3}$ & 2.4 $\pm$ 0.2  & 0.32 $\pm$ 0.16 & -1.27 \\
$K^{0}_{e 3}$ &    0    & 0.46 $\pm$ 0.08 & -0.32 \\
$K^{+}_{\mu 3}$ & 2.4 $\pm$ 0.2  & 0.006 $\pm$ 0.16 & $0.0 \pm 1.0$ [*] \\
$K^{0}_{\mu 3}$ &  0  & 0.15 $\pm$ 0.08 &  $1.7 \pm 1.0$ [*]  \\
\hline 
\end{tabular}
\caption{Summary of isospin-breaking factors from Ref.~\cite{CKNRT}, in units 
of $10^{-2}$; the entries with [*] are from Ref.~\cite{ginsberg}.}
\label{tab:iso-brk}
\end{table}

\begin{figure}[t]
\begin{center}
$$
\includegraphics[width=8 cm]{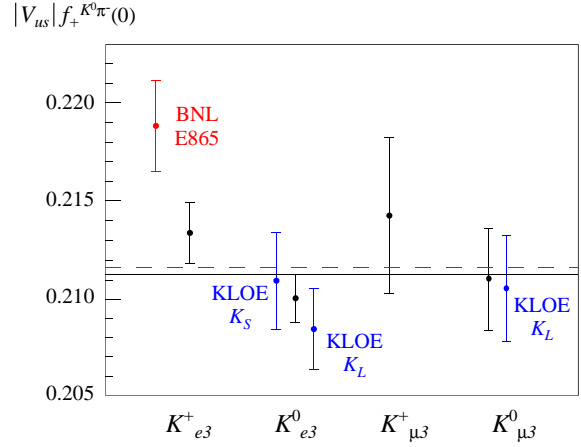}
$$
\caption{ $|V_{us}| \cdot f_{+}^{K^0 \pi^-} (0)$ from the
four $K_{\ell 3}$ modes, including the very recent result 
from BNL-E865 \cite{E865} and preliminary results from KLOE \cite{KLOE};
the (black) points without labels correspond to the
old published results averaged by PDG \cite{pdg2002}.
The full (dashed) horizontal line denotes the average 
without (with) the new data.}
\label{fig:average}
\end{center}
\end{figure}

Applying this master formula to the presently available data 
leads to the plot in Fig.~\ref{fig:average}, which provides
an update of similar analyses presented in Ref.~\cite{CKMBook,CLC}.
The first point to be noted is that the new generation of 
experiments already produced single measurements which 
compete with the PDG values. Performing a naive 
average of all the points in  Fig.~\ref{fig:average}
leads to an error of  $|V_{us}| \cdot f_{+}^{K^0 \pi^-} (0)$
around $0.3\%$, which would be negligible with respect
to the theoretical uncertainty in $f_{+}^{K^0 \pi^-} (0)$ (point n.~4 in the list).
However, we cannot perform a naive average of all the points
in  Fig.~\ref{fig:average} given their internal consistency:
if we include a scale factor following the usual PDG 
procedure \cite{pdg2002}, the error goes up to about 0.8\%,
which is worse than what is obtained without the new results. 
The central value of the average moves by 
less than $0.3\%$ with the inclusion of the new points,
this is why we stated in the previous section that the 
overall picture is essentially unchanged with 
respect to Ref.~\cite{CKMBook}.

As can be seen from Fig.~\ref{fig:average}, the only point 
which is badly consistent with the others is the new  $B(K^{+}_{e3})$
measurement by BNL-E865. This new result differ by $2.3\sigma$ 
with the average of older measurement of the same channel, 
and by more than $3\sigma$ with the average of the neutral modes 
(once theoretical isospin-breaking corrections are applied). 
Given this situation,
it is clear that new independent measurements of $B(K^{+}_{e3})$
--- soon expected by KLOE and NA48~\cite{NA48} -- are particularly 
interesting. If the BNL-E865 result is confirmed, it means that 
isospin-breaking corrections have been badly underestimated 
in Ref.~\cite{CKNRT} and the extraction 
of $|V_{us}|$ from $K_{\ell 3}$ decays is more
complicated than expected. If charged and neutral modes
turn out to be compatible, we would become more confident 
about the theoretical treatment of $K_{\ell 3}$ decays and 
we could hope to reach, in a short time, an overall error 
on $|V_{us}|$ substantially below $1\%$.

As far as the theoretical estimate of $SU(3)$-breaking 
is concerned, an interesting new development is provided 
by the work of Bijnens and Talavera \cite{BT}. They have  
pointed out that, within CHPT, the local $SU(3)$-breaking
contribution to $f_+(0)$ of $\cO(p^6)$ (i.e. the 
leading local contribution), can be unambiguously predicted in terms 
of the the first two derivatives of $f_0(t)$ (which in 
principle are experimentally accessible). In particular,
the slopes $\lambda_0$ and  $\lambda^\prime_0$, defined by  
$$
 f_0(t) =f_+(0)\left[ 1 + \lambda_0 \frac{t}{m_\pi^2}
+  \lambda^\prime_0 \frac{t^2}{m_\pi^4} \right]~, \qquad  t=(p_K -p_\pi)^2, 
$$
should be measured with absolute
errors of $10^{-3}~(\lambda_0)$ and  $10^{-4}~(\lambda^\prime_0)$
in order to reach a prediction of  $f_+(0)$
at the $1\%$ level. Although very challeneging, this goal
is not impossible for high-statistics experiments
such as KLOE and NA48.\footnote{~Note that this measurements could be 
performed either in $K^+_{\mu 3}$ or in  $K^0_{\mu 3}$
channels, since the relative isospin-breaking 
corrections are known.} An interesting complementary 
approach to estimate the amount of $SU(3)$-breaking 
in  $f_+(0)$ is provided by Lattice QCD. 
Unfortunately, at present none of this new techniques 
can lead to a numerical prediction, and   
the most reliable figure is still represented 
by the Leutwyler-Roos result: $f_{+}^{K^0 \pi^-} (0)=0.961 \pm 0.008$~\cite{LeuRo}.
Combining it with $|V_{us}| \cdot f_{+}^{K^0 \pi^-} (0)  =  0.2115 \pm 0.0015$,
obtained from the published data on $B(K^{+}_{e3})$ and  $B(K^{0}_{e3})$ only,
one finds \cite{Vincenzo}
\bea
|V_{us}|_{K_{\ell 3}} &=&  0.2201 \pm 0.0016_{\rm exp}  \pm 0.0018_{{\rm th}(f_+)} 
\nonumber \\
&=&  0.2201 \pm 0.0024~.
\label{eq:Vus_fin}
\eea

\subsection*{Other determinations of $|V_{us}|$}
Alternative strategies to determine $|V_{us}|$
are offered by $\tau$-lepton and Hyperon decays

\noindent
$\underline{\mbox{Tau decays}}$.
The novel strategy  to determine $V_{us}$  via $\tau$ decays, 
proposed in Ref.~\cite{pich02}
and illustrated in this workshop by Jamin \cite{Jamin},
relies on the fact that, using the OPE, we can express theoretically 
the hadronic width of the $\tau$ lepton and the appropriate moments
 --- for both Cabibbo-allowed ($R_{\tau,V+A}^{kl}$)
and Cabibbo-suppressed ($R_{\tau,S}^{kl}$) transitions --- in terms of 
strange-quark mass  and CKM matrix elements. 
Originally, this feature has been exploited to determine
$m_s$ using $|V_{us}|$ as input.  The authors of
Ref.~\cite{pich02} have inverted this line of reasoning: they have
employed the range $m_s (2 \, {\rm GeV}) = 105 \pm 20$ MeV,   
derived from other observables, to determine $|V_{us}|$ 
from hadronic $\tau$ decays.
Using the lower moments only ($k=l=0$) they obtained 
\bea
|V_{us}|_{\tau}  &=& 0.2173 \pm 0.0044_{\rm exp} \pm 0.0009_{\rm th} \pm 
0.0006_{V_{ud}} \no \\
&=&  0.2173 \pm 0.0045 \ ,
\label{eq:tauvus2}
\eea
where the theoretical error reflects the uncertainty in $m_s$, 
the dependence on $V_{ud}$ correspond to the safe range 
$|V_{ud}|= 0.9739 \pm 0.0025$, and the dominant
experimental error reflects the inputs
$R_{\tau,S} = 0.1625 \pm 0.0066$ and $R_{\tau,V+A} = 3.480 \pm
0.014$~\cite{davier02}.  
A reduction in the uncertainty of $R_{\tau,S}$ by a factor of two,
which should easily be reached at $B$ factories,
would make this extraction of $|V_{us}|$ competitive with the one
based on $K_{e 3}$ decays. As already stressed in Ref.~\cite{CKMBook},
in this perspective it would be highly
desirable also to estimate the systematic uncertainty of the method 
(e.g. extracting $V_{us}$ from higher $R_{\tau,S}^{kl}$ moments, 
and obtaining additional constraints on the $m_s$ range). 
Future precise measurements of $\tau$ hadronic moments,
with a good flavor tag,
could allow to reach this goal.

\medskip

\noindent
$\underline{\mbox{Hyperon semileptonic decays}}$. 
A new analysis of  Hyperon semileptonic decays
has recently been presented in Ref.~\cite{Cabibbo}. 
On general grounds, these processes 
are not so clean as $K_{\ell 3}$ decays since: 
i) the hadronic matrix elements of the axial current
(not protected by the Ademollo-Gatto theorem)
are also involved, ii) the convergence of the 
chiral expansion is slower and the corresponding 
coefficients are known with less accuracy. As shown
in Ref.~\cite{Cabibbo}, the first problem can
be circumvented by fitting the ratio of axial over 
vector current at zero momentum transfer ($g_1/f_1$) 
from data (similarly to what is done for the extraction
of $|V_{ud}|$ from the neutron beta decay).
By doing so, and neglecting possible 
$SU(3)$ and isospin-breaking breaking terms in $f_1(0)$ 
due to quark masses, the authors of Ref.~\cite{Cabibbo} obtain 
\be
|V_{us}|_{\rm Hyp} =  0.2250 \pm 0.0027_{\rm exp}~,
\label{eq:Hypvus}
\ee
where the average is dominated by the two values 
from $\Lambda$ ($0.2224 \pm 0.0034$)
and $\Sigma^-$ ($0.2282 \pm 0.0049$)
semileptonic decays. The fact that the error in 
(\ref{eq:Hypvus}) is very close to the final error
in (\ref{eq:Vus_fin}) and it is in better agreement 
with CKM unitarity is rather stimulating. However, 
we stress that the comparison between (\ref{eq:Hypvus}) and 
the final error  in (\ref{eq:Vus_fin}) is not appropriate, 
since 
the latter {\em does include} an estimate 
of the theoretical uncertainty due to light-quark masses. 
The calculation of  $SU(3)$-breaking effects 
in the matrix elements of the vector current at 
zero momentum transfer is more difficult in the baryonic 
sector than in meson one, and indeed the existing
estimates are affected by sizable uncertainties 
(see e.g. Ref.~\cite{Luty}).\footnote{~Note that in the
case of the $f_+(0)$ the leading non-local term of 
$\cO(p^4)$ is known to excellent accuracy and the first 
ambiguities arises at the two-loop level, 
while in the baryonic sector there are sizable 
ambiguities already at the one-loop level \cite{Luty}.}
For this reason, we believe that the error in 
(\ref{eq:Hypvus}) cannot be considered as very 
conservative. However, we fully agree with the 
statement of Ref.~\cite{Cabibbo} that the situation 
might improve in the future with help of Lattice QCD.

\section{Status of $|V_{ud}|$ and CKM Unitarity}
The situation of $|V_{ud}|$ has not changed 
since the publication Ref.~\cite{CKMBook}. 
As stressed by Abele \cite{Abele} at this workshop,
the nine independent measurement of SFT in different 
nuclei show a remarkable internal consistency once 
the appropriate universal and structure-dependent 
radiative corrections are included. The latter have been 
recently re-analyzed in Ref.~\cite{Towner_new},
confirming (and thus strengthening the confidence in) 
the older analyses, which leads to 
the global average \cite{Hardy}
\begin{eqnarray}
 |V_{ud}|_{\rm SFT} &=& 0.9740 \pm 0.0005\ .
 \label{eq:vudsft}
\end{eqnarray}

The internal consistency of the neutron beta decay
determination of  $|V_{ud}|$ is also rather good 
once we restrict the attention to recent experiments
with a high degree of polarization to measure $g_A/g_V$
(which represent the dominant source of uncertainty).
Their average leads to \cite{CKMBook}
\begin{equation}
  |V_{ud}|_{n_\beta} = 0.9731 \pm 0.0015~,
\label{eq:nbfin}
\end{equation}
which is expected to improve substantially 
in the near future thanks to the upgrade of the 
PERKEO experiment in Heidelberg \cite{Abele}.

The third and completely independent approach to $|V_{ud}|$, 
namely the determination via the $\beta$-decay of the 
charged pion, appears to be very promising in the
long term due to the excellent theoretical accuracy
of the corresponding decay amplitude \cite{cknp02}.
The present experimental precision for the tiny 
branching ratio of this transition does not allow yet to compete
with SFT and $n_\beta$ determinations; however, the situation 
is improving thanks to the PIBETA experiment at PSI \cite{PIBETA}.
The preliminary result of the PIBETA Collaboration \cite{Pocanic},
$$
B(\pi^+\to \pi^0 e^+\nu) = (1.044 \pm 0.007_{\rm stat.} \pm 0.009_{\rm syst.}) \times  10^{-8},
$$
combined with the theoretical analysis of Ref.~\cite{cknp02}, leads to 
\begin{eqnarray}
|V_{ud}|_{\pi_{e3}} &=& 0.9765 \pm 0.0056_{\rm exp} \pm 0.0005_{\rm th}
\nonumber \\
         &=& 0.9765 \pm 0.0056~,
\end{eqnarray}
where the error should be reduced by about a factor of 3 at the 
end of the experiment.

\subsection*{CKM unitarity}
The two measurements of $|V_{ud}|$ from SFT and nuclear beta decay, 
reported in  Eqs.~(\ref{eq:vudsft}) and (\ref{eq:nbfin}) respectively,
are perfectly compatible. Combining them in quadrature one obtains 
\be
|V_{ud}| = 0.9739 \pm 0.0005~,
\label{eq:Vud_fin}
\ee
a result which is not modified 
by the inclusion in the average of the present $\pi_{e 3}$ data.
The compatibility of SFT and nuclear beta decay results 
is clearly an important consistency check of Eq.~(\ref{eq:Vud_fin}).
However, it should also be stressed that the theoretical uncertainty 
of inner radiative corrections (which contribute at the level of 
$\pm 0.04\% $) can be considered to a good extent 
a common systematic error for both determinations. 
Thus the uncertainty quoted in Eq.~(\ref{eq:Vud_fin}) is mainly 
of theoretical nature and should be taken with some care.
Using the unitarity relation (\ref{eq:unitarity}) we can translate 
Eq.~(\ref{eq:Vud_fin}) into a prediction for $|V_{us}|$:
\be
|V_{us}|_{\rm unit.} = 0.2269  \pm 0.0021~,
\label{eq:Vus_unit}
\ee
to be compared with the direct determination in Eq.~(\ref{eq:Vus_fin}). 

As already pointed out in Ref.~\cite{CKMBook},
the $2.2\sigma$ discrepancy between these two determinations could
be attributed to: i) an underestimate of theoretical and, more general,
systematic errors; ii) an unlikely statistical fluctuation; 
iii) the existence of new degrees of freedoms which spoil the unitarity of the CKM matrix. 
Since theoretical errors provide a large fraction of the total
uncertainty in both cases, at present the solution i), 
or at least a combination of i) and ii), appears to be the most likely scenario. 
As discussed in Section~2, the situation of $K_{\ell 3}$
decays is in rapid evolution: with help of 
new data and new theoretical estimates of $SU(3)$-breaking 
effects in these channels, we should be 
able to shed new light on these three scenarios 
in the near future. For the time being, if 
we are interested in a conservative estimate 
of the Cabbibbo angle to be used in different frameworks
(e.g. global CKM fits),  the best we can do is to
treat the two determinations in (\ref{eq:Vus_fin}) and (\ref{eq:Vus_unit})
on equal footing and to introduce
an appropriate  scale factor. Following this procedure, 
we confirm the estimate of $|V_{us}|$
presented in \cite{CKMBook}, namely
\be
|V_{us}|_{{\rm unit.}+K_{\ell 3}} = 0.2240 \pm 0.0034~.
\label{eq:Vus_global}
\ee

\section*{Acknowledgments} 
I wish to thank Mario Antonelli, Vincenzo Cirigliano, Gilberto Colangelo, 
Paolo Franzini, Claudio Gatti, Gabriel Lopez-Castro, Philip Ratcliffe,
Barbara Sciascia, Tommaso Spadaro and Julia Thompson for 
useful comments and/or discussions. This work is  partially supported 
by IHP-RTN, EC contract No. HPRN-CT-2002-00311 (EURIDICE).

\end{document}